# Enhancing Big Data in the Social Sciences with Crowdsourcing: Data Augmentation Practices, Techniques, and Opportunities

**Nathaniel D. Porter**

Ph.D. Candidate
Department of Sociology
Pennsylvania State University
ndp135@psu.edu

**Ashton M. Verdery**

Assistant Professor
Department of Sociology
Pennsylvania State University
amv5430@psu.edu

**S. Michael Gaddis**

Assistant Professor
Department of Sociology
University of California – Los Angeles
mgaddis@soc.ucla.edu

**May 2017 Draft**
Do not cite without authors' permission

**Abstract**

The importance of big data is a contested topic among social scientists. Proponents claim it will fuel a research revolution, but skeptics challenge it as unreliably measured and decontextualized, with limited utility for accurately answering social science research questions. We argue that social scientists need effective tools to quantify big data's measurement error and expand the contextual information associated with it. Standard research efforts in many fields already pursue these goals through data augmentation, the systematic assessment of measurement against known quantities and expansion of extant data by adding new information. Traditionally, these tasks are accomplished using trained research assistants or specialized algorithms. However, such approaches may not be scalable to big data or appease its skeptics. We consider a third alternative that may increase the validity and value of big data: data augmentation with online crowdsourcing. We present three empirical cases to illustrate the strengths and limits of crowdsourcing for academic research, with a particular eye to how they can be applied to data augmentation tasks that will accelerate acceptance of big data among social scientists. The cases use Amazon Mechanical Turk to (1) verify automated coding of the academic discipline of dissertation committee members, (2) link online product pages to a book database, and (3) gather data on mental health resources at colleges. In light of these cases, we consider the costs and benefits of augmenting big data with crowdsourcing marketplaces and provide guidelines on best practices. We also offer a standardized reporting template that will enhance reproducibility.

**INTRODUCTION**

Big data and computational approaches present a potential paradigm shift in the social sciences, particularly since they allow for measuring human behaviors that cannot be observed with survey research (Lazer et al. 2009; Moran et al. 2014). In fact, the transformative potential of big data for the social sciences has been compared to how "the invention of the telescope revolutionized the study of the heavens" (Watts 2012:266). However, social scientists have been slow to embrace big data. One reason why is "the need for advanced technical training to collect, store, manipulate, analyze, and validate massive quantitates of semistructured data" (Golder and Macy 2014:144), training that remains nascent in many fields. But there are deeper, more fundamental constraints on the acceptance of big data among social scientists.

Despite its promise, big data's perceived limitations cast uncertainty on its applicability in the social sciences. Computer, information, and physical scientists have rapidly embraced big data because the information it makes available is unprecedented in those fields. Typical taxonomic efforts from computer scientists and others to delineate big data from traditional forms of data focus on these novel characteristics in what is called the "three Vs" framework (Hitzler and Janowicz 2013; Yin and Kaynak 2015): volume (or amount of data), velocity (or speed of data release), and variety (or data on rarely recorded activities). Volume, velocity, and variety are what make big data compelling and useful in a diverse array of fields. However, social scientists are concerned with two other Vs: validity[1] and value (Hitzler and Janowicz 2013; Monroe 2013). These additional Vs, which indicate authenticity or truth (validity) and what we can do with and get out of the data (value), are often lacking in big data research (Monroe 2013; Yin and Kaynak 2015). Characteristic of social science skepticism around big data are concerns that "[t]he reliability, statistical validity and generalizability of new forms of data are not well understood. This means that the validity of research based on such data may be open to question" (Entwisle and Elias 2013:1). Put bluntly, big data do not come from a heavily theorized and well planned scientific research project, which, at a minimum, creates discomfort among social scientists (Lazer and Radford 2017).

---

[1] Other computational and information scientists refer to the 5 Vs of big data as including volume, velocity, variety, value, and veracity. Political scientists swap value and veracity for 'vinculation' (to bind together in a relationship) and validity, quipping: "[t]here are as many 'fourth Vs of B big data' as there are 'fifth Beatles'" (Monroe 2013:1). We stray slightly from this jargon and refer to veracity as validity in the rest of this paper to more closely match the language of social science methods.

Without clear approaches to quantify and increase the validity and value of big data, we believe social science skepticism of big data will remain high. Researchers need to be convinced of the validity and value of big data, while simultaneously not adding to the cost of using big data, all of which we suggest can be accomplished through data augmentation. We define data augmentation as the process of (a) systematic assessment of measurement against known quantities or (b) expansion of extant data by adding new information.

Data augmentation is a standard technique throughout the social sciences that can assume a manual or automated approach. Traditionally, these tasks are accomplished using trained research assistants (manual) or specialized algorithms (automated) to detect erroneously coded data (validity) or append existing data sources with new material (value). An example of a manually augmented big data project is a study of posts made by high-schoolers on the Twitter social media platform that mention bullying. In this study, the authors used two human coders to classify whether each post that mentioned bullying (or bullied, or bully, etc.) was an actual report of adolescent bullying or whether it represented some other use of the relevant terms (Bellmore et al. 2013). In this case, the authors used data augmentation to increase validity. An example of automated data augmentation used to increase value is a well-known experiment on the social media platform Facebook (Kramer, Guillory, and Hancock 2014). In this experiment, the authors examined how respondents' purported emotions changed after being shown more purportedly positive or negative posts from friends, where emotions and their associated positivity or negativity was assessed by applying a sentiment analysis method to the words used in posts. Sentiment analysis, in this case, serves as an automated way to gain additional information about big data (the posts), augmenting its value for research purposes. Of course, there are many more examples of both manual and automated approaches to data augmentation to add either validity or value or both (e.g., Maldonado et al. 2015; Bail 2016).

Unfortunately, data augmentation can be challenging to implement at the scale required for big data projects in a way that addresses social science skepticism. The manual data augmentation in the aforementioned study of bullying, for instance, was only feasible because the researchers examined a manageable number of messages (N=7,321). Automated data augmentation approaches, such as sentiment analysis, are also difficult to implement without advanced training and may themselves be of questionable validity. For instance, the automated augmentation used in the Facebook experiment discussed above has been criticized by social scientists for being of unknown, and potentially low, validity (Panger 2016). Of course, the validity of automated data augmentation approaches can be assessed and potentially improved through manual data augmentation, as is becoming more commonplace in big data projects through procedures such as supervised machine learning (Bail 2014), but the size and complexity of most big data would require a great deal of time and expense for knowledgeable trained coders (such as graduate assistants) to check.

In this paper, we argue that online crowdsourcing platforms can complement both manual and automated approaches to data augmentation, increasing the validity and value of big data in the social sciences at a low cost to researchers. We show that such tools are underused for non-experimental designs in the social sciences and that workers on these platforms can rapidly and inexpensively verify automated coding, find errors in embedded metadata, and resolve missing data in many cases. We build this case in five steps: (1) review the use and perceived limitations of big data in the social sciences, (2) describe the online crowdsourcing process and its documented strengths and limitations as a platform for academic research, (3) investigate current practices in academic use of the largest online crowdsourcing platform, (4) conduct three case studies implementing online crowdsourcing to enhance ongoing sociological research and test the utility of crowdsourcing across different circumstances, and (5) draw on all of the above, as well as experiments embedded within the case studies, to produce evidence-based recommendations on when and how to implement online crowdsourcing to augment big data for best results. Finally, in light of the inconsistent and frequently incomplete reporting of online crowdsourcing procedures, we provide a recommended reporting template for online crowdsourcing as an academic data augmentation platform. We believe that this paper offers a clear roadmap for social scientists to begin incorporating more big data into their research designs, and we conclude by reflecting on the strengths and limits of online crowdsourcing approaches to data augmentation for these purposes.

**Big Data Skepticism in the Social Sciences**

Myriad actors such as corporations, governments, scientists, and even sports teams have embraced big data (Lohr 2012; Mayer-Schönberger and Cukier 2013; Murdoch and Detsky 2013) but adoption has been slow thus far in the social sciences (Lazer and Radford 2017). To understand how social science adoption of big data compares to its use in other fields, we searched Thompson-Reuters' Web of Science database in April 2017 for academic articles with the phrase "big data" (with quotes, not case-sensitive) appearing in the title, abstract, or keywords. The phrase gained its contemporary meaning in 2004; for a number of years thereafter, only a handful of isolated articles drew on the idea. Figure 1 shows the time series of papers about "big data" from 2009-2016, both overall and by some key fields. Beginning around 2011, overall use of big data began to increase exponentially. The increase has not been even across fields, however, as growth has been concentrated in computer science and other computationally intensive fields like engineering. By contrast, social science use remains minimal, with, for example, only 70 publications categorized as sociology between 2004 and 2016 (1.37% of all publications listing big data).[2]

---

[2] Thompson Reuters' classification scheme for research areas may not correctly identify sociology articles, or sociologists may be publishing big data articles in non-sociology journals. We do not feel that these possibilities restrict our general conclusions, because, in either case, sociologists will experience less exposure to big data articles. A related concern is that other fields simply produce more research than the social sciences, thereby accounting for the small role of the social sciences in big data research. However, research into article counts by discipline do not indicate the levels of disparity seen in Figure 1. For instance, Jaffe (2014) shows that the social sciences and psychology produced approximately 150,000 articles in 2011, whereas engineering produced approximately 250,000.

**Figure 1. Numbers of articles with topic "Big Data" overall and matching select fields, 2009-2016**

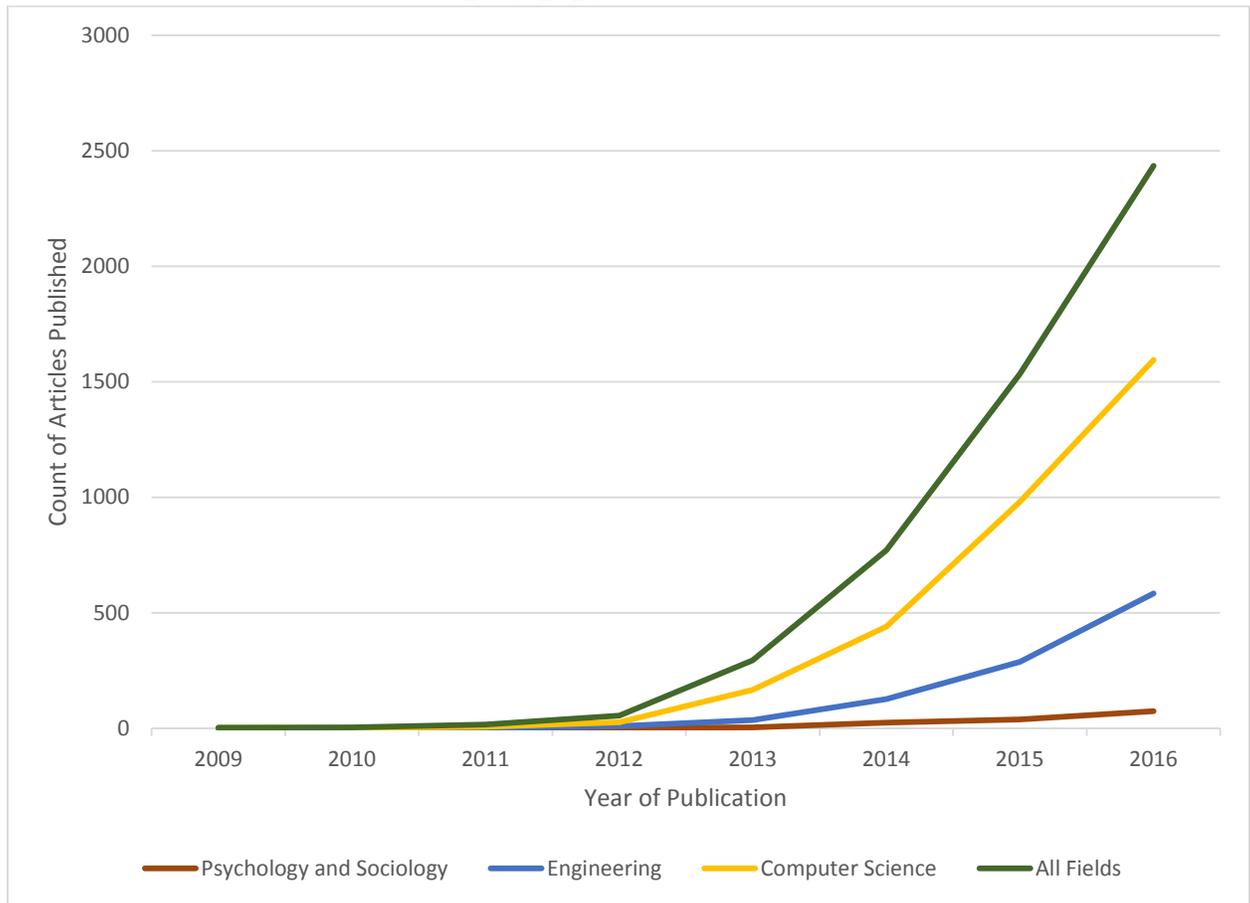

Source: Thompson Reuters Web of Science, search conducted 2017-04-18.

The literature indicates that the primary reason social scientists are making relatively rare contributions to big data research is that these fields hold deep skepticism about big data deriving from the fact that it is not designed for academic research (Lazer and Radford 2017). Even those optimistic about the promise of big data critique its validity and value, including its lack of standardized reporting (K. Lewis 2015), poor measurement (Diesner 2015), decontextualization (Bail 2014), and tendency toward "big data hubris" (Lazer et al. 2014) that ignores threats to validity (Adams and Brückner 2015; Park and Macy 2015). Generalizability is another concern; most big data studies do not proceed with a clearly conceptualized population to which inference can be made (Boyd and Crawford 2012; Clifford, Jewell, and Waggoner 2015; Lazer and Radford 2017; Weinberg, Freese, and McElhattan 2014). Disciplinary divisions in computational skills (McFarland, Lewis, and Goldberg 2015, Leetaru 2014) and epistemology pose additional challenges (Wagner-Pacifici, Mohr, and Breiger 2015), as do divides between industry and academic research (Boyd and Crawford 2012). However, federal funders and several universities have funded a wide range of new training programs and other undertakings at the nexus of big data and the social sciences that may, over time, alleviate these pressures (e.g., http://bdss.psu.edu/, http://dsi.ucdavis.edu/, http://www.fragilefamilieschallenge.org/).

The broad range of concerns about big data from social scientists has led to a number of reflections on what steps can be taken to address this skepticism. However, our reading of the literature indicates that these reflections have focused more on the issues of generalizability than other, equally important concerns. For instance, in their review article, Lazer and Radford (2017) list the vulnerabilities of big data research in sociology. The primary listing – indeed "[t]he core issue…" – is generalizability – "… *who* and *what* get represented" (Lazer and Radford 2017: 13, italics sic). While these authors do acknowledge validity and value concerns, they are given only marginal discussion. We feel that this is an oversight that reveals a fundamental gap between what researchers worry about with big data and what is being done to address those worries.

In general, the primary means of assessing and increasing the validity and value of data in the social sciences is undertaken through what we refer to as data augmentation. As reviewed above, there are both manual and automated approaches to data augmentation, but neither is likely to be sufficient to rise to the scale of the problems posed by big data and address social science skepticism about it. Instead, we focus on a third option that can enhance both automated and manual approaches to data augmentation: using online crowdsourcing marketplaces such as Amazon Mechanical Turk (MTurk). Online crowdsourcing is less technically demanding than automated approaches and can provide supplemental evidence of accuracy based on user judgment or augmented comparison with outside sources or both. Compared to common manual approaches, MTurk is nimbler and less costly, allowing increased scale of augmented analysis. Compared to purely automated approaches or even blended approaches like supervised machine learning, online crowdsourcing through MTurk has the ability to produce well-understood measures of validity like inter-rater reliability or to merge data with sources that are not amenable to automated discovery, as well as retaining the reassuring feature that actual people have examined the coding. While some social scientists are using MTurk for research (Flores 2016; Gaddis 2017), we argue that formalizing this approach to data augmentation will expedite the widespread acceptance of big data in the social sciences and overcome barriers to its application. In the next section, we review MTurk as a promising research platform that we argue allows researchers to undertake big data augmentation at scale more simply, quickly, and cheaply than data augmentation through traditional automated or manual approaches.

**MTurk as a Research Platform**

The name "Mechanical Turk" is derived from the 18$^{th}$ century chess-playing "machine." The original Mechanical Turk consisted of a complex cabinet of gears with a magnetic chessboard on top and a model of a human similar to a mannequin dressed in Turkish robes with a turban. Human chess players could play against the "machine" and would often lose. The Mechanical Turk toured Europe and the United States throughout the late 18$^{th}$ and early 19$^{th}$ centuries. However, the Mechanical Turk was a hoax as it was not an automated machine but rather an elaborate fake with a man inside playing the actual chess game (Levitt 2006; Standage 2004).

Thus, Amazon named their own version after the original Mechanical Turk to indicate that humans can still do things that computers cannot. Amazon's MTurk is an online crowdsourcing marketplace that brokers what MTurk parlance refers to as Human Intelligence Tasks (HITs) between requesters and workers[3]. The idea of a HIT is described succinctly by Amazon:

> Amazon Mechanical Turk is based on the idea that there are still many things that human beings can do much more effectively than computers, such as identifying objects in a photo or video, performing data de-duplication, transcribing audio recordings, or researching data details. Traditionally, tasks like this have been accomplished by hiring a large temporary workforce (which is time consuming, expensive, and difficult to scale) or have gone undone.[4]

Anyone eligible for employment in the U.S. or India can work on MTurk, although task completion requires reliable internet access. U.S.-based MTurk workers are typically younger, more educated, wealthier, more technologically savvy, and less racially diverse than average Americans (Berinsky, Huber, and Lenz 2012; Krupnikov and Levine 2014; Paolacci and Chandler 2014). As such, many worry that samples drawn from MTurk are less representative than population based surveys (Berinsky, Huber, and Lenz 2012), though not as fraught as convenience samples (Buhrmester, Kwang, and Gosling 2011).

However, when considering MTurk as a big data augmentation platform, as we propose, rather than a population to sample and survey, we argue that work quality matters more than worker representativeness. MTurk workers tend to pass screening tests at high rates (Berinsky, Huber, and Lenz 2012) with high reliability between (Behrend et al. 2011) and within workers

---

[3] Similar sites such as MicroWorkers and CloudFactory serve more specialized clienteles, but MTurk is the oldest and largest such site, with more than 500,000 registered workers (Kuek et al. 2015).
[4] https://www.MTurk.com/MTurk/help?helpPage=overview. Accessed January 6, 2016.

(A. R. Lewis et al. 2015). Recruiting workers for data augmentation tasks through MTurk has three major limitations. First, workers lack specialized area knowledge; second, they cannot access restricted information (e.g. workers cannot download most academic journal articles); and third, MTurk compensation is based on task completion, not time, which presents challenges for fielding complex, judgment based tasks (Goodman, Cryder, and Cheema 2013; Krupnikov and Levine 2014). We return to these ideas below. For now, it is worth noting that these limitations mean that crowdsourced tasks are most appropriate for data augmentation when they can be broken into *concise* and *unambiguous* chunks using *non-confidential* information.

**MTurk in the Academy**

MTurk is popular with academic researchers; a recent Pew Research Center report (Hitlin 2016) found that academics posted the plurality (36%) of all HIT groups during one week. Academics have hailed MTurk's low costs and rapid results, and even expressed cautious optimism about it as a survey platform (Horton, Rand, and Zeckhauser 2011; Weinberg, Freese, and McElhattan 2014). Its feasibility and reliability for big data augmentation, however, remains unexplored.

To better understand how academics use MTurk, especially for data augmentation, as well as how they report on such use, we conducted a content analysis of a random with-replacement sample of 100 articles from Web of Science matching the topic search "mechanical turk" and published between 2011 and May 2016. The search, performed May 23, 2016, returned 767 total records. We removed eight false matches, one poster, and three papers we could not find, yielding a final sample size of 88 articles (80 unique; statistics below are weighted for replacement sampling). In the online supplement, we provide metadata about these articles. We address three questions in this content analysis: a) who uses MTurk for academic purposes, b) what is it used for, and c) what details are reported about the use of the platform.

Over half (61%) of the papers we examined were in psychology and related fields (psychiatry, social psychology, and cognitive science), followed by business and organizational fields (10%), computer science and engineering (9%), and (non-mental) health fields (6%). The remaining 13% of articles came from many disciplines, including law, linguistics, anthropology, political science, and sociology. Article counts grew steadily from MTurk's founding in 2011 through 2015, the last full year in our data. In general, these articles are cited frequently, with Web of Science's citation counts indicating an average of 16 citations (median 8) for articles at least two years post-publication. These levels compare favorably to general article citation counts across many fields, where citation counts often average one per year or less (Thompson Reuters 2010).

We are also interested in what researchers use MTurk for, specifically how often it is used for data augmentation. Table 1 reports on the types of tasks academic researchers assign to MTurk workers. Because of psychology's disproportionate use of MTurk, we disaggregate results by whether the article was in a psychological field. Most papers used MTurk to field surveys (64%), but data augmentation comprised the second most common category (59%). In our sample, non-psychological studies (76%) used MTurk more often than psychological studies (48%) for data augmentation. Of the studies involving data augmentation, workers are more commonly asked to augment provided data, but never asked to collect publicly available data from the web. We view this latter use as particularly promising avenue for big data augmentation with MTurk.

**Table 1. Worker tasks in 100 Articles Matching Topic "Mechanical Turk" in Web of Science**

|  | Psychology (N=54) | Other Fields (N=34) | Total (N=88) |
|---|---|---|---|
| *Take Surveys** | 80% | 38% | 64% |
| *Pilot Studies* | 28% | 44% | 34% |
| *Experimental Designs* | 48% | 53% | 50% |
| *Data Augmentation* | 48% | 76% | 59% |
|     Verify/Replicate Other Data | 41% | 59% | 48% |
|     Elaborate on Data Provided by Researcher | 0% | 21% | 8% |
|     Code Factual Data Provided by Researcher* | 0% | 21% | 8% |
|     Collect Publicly Available Data from Web | 0% | 0% | 0% |

Notes:
Many studies ask workers to complete multiple tasks, so major categories percentages do not add to 100%.
* Two-tailed F test between psychology and other fields significant (p<.001).

    Another question of interest is how academic researchers report on their use of MTurk as a data augmentation platform. Although researchers use MTurk for data augmentation, we found gaps in reporting standards that may impair the validity, value and replicability of MTurk as a data augmentation tool. Nearly every article we examined (92%) described data collection procedures like HIT content in detail, and most (80%) included at least basic summaries of worker demographics. However, few articles we examined reported required worker qualifications, criteria for work rejection, or validation criteria. Only 16% met what we define as basic reporting standards across three key areas for peer evaluation and replicability: a) a detailed description of the HITs and process, b) information on worker qualifications, acceptance criteria and pay, and c) descriptive statistics, multivariate analysis, or formal validity checks.

    The results of our content analysis highlight that academic use of MTurk remains concentrated in psychological fields, and for experimental studies, piloting, and surveys. In contrast to this typical use, we advocate that researchers expand their use of MTurk for

augmenting big data studies to address concerns about validity and value. We found that researchers are beginning to do this, but they do not offer enough detail on the process for it to be formally evaluated. To this end, the remainder of this article examines three case studies and focuses on developing clear, evidence-based guidelines for best practices on when and how researchers can augment data with MTurk and report on doing so.

**Case Studies**

We now present three case studies that apply MTurk to diverse sociological subfields to augment big data (cases 1 and 2) or test MTurk's data augmentation capacities against known benchmarks from ongoing sociological data collection (case 3). These cases allow us to compare MTurk to other data augmentation approaches, both automated and manual. For cases 1 and 3, we collected analogous data automatically and manually, enabling validity comparisons. We also embedded design experiments in cases 2 and 3 to test how HIT design and implementation can affect cost, quality, and worker experience. Our goal is to develop intuition for the benefits of big data augmentation through online crowdsourcing and how researchers can best move forward with such projects.

We designed all HITs based on past recommendations (Buhrmester, Kwang, and Gosling 2011; Paolacci, Chandler, and Ipeirotis 2010; Berinsky, Huber, and Lenz 2012) and revised them according to common worker concerns voiced in online MTurk forums (e.g., http://www.turkernation.com) and our own piloting. We collected all data between October 2015 and July 2016. The online supplement provides full versions of instruments and de-identified results.

*Study 1: Academic Affiliation – Overview and Methods*

Our first case shows how MTurk can enhance the validity of big data. It is part of a larger project on the role of interdisciplinary dissertation committees in knowledge production . The original project used an algorithm to code the academic field of faculty based on their roles in doctoral committees. For instance, if a faculty member chaired committees in one field and was a member of committees in another, the algorithm assigned them to the field in which they chaired. Most cases were less clear cut, however, and required more complex assignment rules reviewed in greater depth in the original paper. Such algorithmic assignment indicated a surprising amount (56%) of interdisciplinary dissertation committees. The credence given to these prevalence statistics, however, hinges on the accuracy of the automated coding. This represents a classic concern voiced by social science skeptics about automated augmentation of big data. For instance, compare the critique of sentiment analysis in the aforementioned Facebook experiment (Panger 2015; Kramer, Guillory, and Hancock 2014) or concerns about search term inclusion in Google Flu (Lazer et al. 2014; Ginsberg et al. 2009). Manually verifying a sample – manual data augmentation – represents one way to check result accuracy, however, our tests indicated that finding and hand coding the fields of a sample of 2,000 of the 66,901 faculty (3%) would have demanded over 230 hours of trained coder work. This time commitment translates to more than three quarters of a semester of typical graduate research assistant support, assuming a 15 week semester at 20 hours a week.

Rather than training internal coders to verify these results, we tested the data augmentation capabilities of MTurk. We did so by creating three sequential tasks that split the process of validating the algorithmic coding of faculty members' fields into discrete steps. First, we asked workers to find the departmental webpages of a random sample of faculty members using a search link that limited results to the official website of their academic institution (see discussion and appendix for details). This step provided a sample of faculty whose academic field could be externally validated. Second, we asked workers to verify links obtained in task 1 and indicate whether each faculty member was listed in any of the 10 most common department names in the algorithmically coded field. This step helped to ensure that the links for specific faculty were correct. Finally, in the third task, we asked workers to evaluate whether any field on the faculty member's page is associated with the field that was algorithmically assigned. For instance, if a faculty member listed "speech and pathology" as their field and the assigned field is "speech and hearing sciences," we aspire for workers to select that these fields are associated. This step constituted our primary interest, quantifying the validity of the algorithmic coding. We adapted all tasks from MTurk templates using the HTML and JavaScript programming languages, and collected them from separate but potentially overlapping pools of workers within the MTurk interface. A graduate research assistant invested approximately 40 hours in learning and managing this MTurk data collection. In all, we used MTurk data augmentation to check 2,043 automated classifications of faculty member fields, at a total cost of $590 including fees and pilot costs.

*Study 1: Academic Affiliation – Results and Discussion*

Were MTurk workers, operating without substantial oversight or prior training, able to validate the results assigned by algorithm? This case speaks to MTurk's ability to add validity to big data, used here to confirm the automated coding of a large data set and bound rates of coding error. Table 2 summarizes the combined results for Case 1. Workers in the initial HIT successfully located 85% of faculty, mostly on preferred page types (faculty homepage, administrative list, or curriculum vitae). Subsequent workers flagged only 3% of URLs that prior workers submitted as referring to the incorrect person or institution. Of cases with unflagged

URLs, workers identified 94% of faculty members as matching either the field or department we provided, which suggests that the original automated coding of these big data succeeded at a high rate, even allowing for the possibility of substantial worker error. Mean hourly worker pay in this case ranged from $7 to $16 and was higher for workers completing multiple HITs.[5]

**Table 2. Contingency Table of HIT Results for Study 1**

| URL Found | | Field Matched | | Department Matched | |
|---|---|---|---|---|---|
| No | 15.5% | | NA | | NA |
| Yes | 84.5% | Unclear | 3.9% | Bad URL | 5.9% |
| | | | | No | 27.9% |
| | | | | Yes | 66.2% |
| | | Bad URL | 2.0% | Bad URL | 34.3% |
| | | | | No | 22.9% |
| | | | | Yes | 42.9% |
| | | No | 13.1% | Bad URL | 2.2% |
| | | | | No | 44.5% |
| | | | | Yes | 53.3% |
| | | Yes | 80.8% | Bad URL | 0.9% |
| | | | | No | 12.4% |
| | | | | Yes | 86.7% |

This case revealed some important lessons. Early pilots combined all stages (page location, department classification, and field classification) into a single HIT, but we found that workers took longer and gave flagged results more often in such conditions. With later pilots, we found that dividing tasks into the three steps outlined above minimized worker time and let us build in cross-verification tests where subsequent workers verified both the faculty web pages and affiliations provided by earlier workers.

---

[5] We report hourly worker pay as an adjusted minimum hourly rate, as workers are allowed to accept multiple HITs at once, thus deflating uncorrected pay rate calculations in multi-HIT batches. See reporting template in appendix for a complete description of the correction.

*Study 2: Linking to OpenLibrary – Overview and Methods*

Our second case highlights how data augmentation with MTurk can enhance the value of big data. Here, we asked workers to connect data sources (adding value to big data), and we experimentally tested how HIT design may affect work quality. This case builds on a project investigating book co-purchasing patterns connecting cultural groups, operationalized with retailer metadata scraped from the web. Unfortunately, necessary metadata were often incomplete, missing, or of questionable quality. For example, a book written by the founder of one Protestant denomination (Martin Luther) was listed as the top-selling item associated with a completely different denomination. To supplement missing information, we matched 1,055 (58%) books to additional metadata provided by OpenLibrary.org using international standard book numbers (ISBNs), a unique code identifying books. For remaining unmatched books, we tested MTurk's data augmentation capacities by asking workers to search for the books on OpenLibrary. As an experiment to determine means of improving HIT design, we randomly assigned each worker into one of three task variants. The first variant included full instructions with design features to enhance clarity (e.g. highlighting key text); the second used brief instructions but retained design features; while the third included full instructions with minimal formatting. Figures 2-4 provide screen shots of each condition; note that Amazon uses the ${variable name} notation as code to substitute values from input data provided by the requester (code available in supplemental files).

**Figure 2: Experimental Variant 1 for Study 2 (complete)**

**Instructions** *(approximate HIT length 1-3 minutes)*

Find and document the **OpenLibrary page** of the book listed below by searching for the title and/or author **here**.

Optionally, you may click the title below to open the Amazon.com product page for the book in a new window, allowing you to verify you have found the correct book.

Some books will have an OpenLibrary page for the title, but not the specific edition. Others may not have an OpenLibrary page at all. When you have answered all questions, click submit.

- Title: ${title}
- Author: ${author}

Did you locate a book with the correct title and author?
- Yes
- No (submit now)

Copy and paste the URL of the works page for the book and then click submit.

If the URL does not include "/works/" it is the wrong kind of page. Clicking a result on a search page should load the works page. To reach it from a specific edition ("book" page), click the title in the top left corner of the page following "N editions of..."

[ https://openlibrary.org/works/... ]

Optional: Type any further comment/explanation below.

[ ]

*Your work will be automatically rejected if you do not answer each applicable question. All other submissions will be randomly verified.*

*If you have any questions or feedback or feel your work was unfairly rejected, please contact the requester. We hope to earn your trust by being both fair and timely in tasks, payment, and communication.*

**Figure 3: Experimental Variant 2 for Study 2 (brief instructions)**

**Instructions** *(approximate HIT length 1-3 minutes)*

Find the **OpenLibrary page** of the book below (if it exists) using this **link**.

- Title: ${title}
- Author: ${author}

Did you locate the book?
- Yes
- No

URL at openlibrary.org:

[ https://openlibrary.org/works/... ]

Optional explanation:

[ ]

*Your work will be automatically rejected if you do not answer each applicable question. All other submissions will be randomly verified.*

*If you have any questions or feedback or feel your work was unfairly rejected, please contact the requester. We hope to earn your trust by being both fair and timely in tasks, payment, and communication.*

**Figure 4: Experimental Variant 3 for Study 2 (plain design)**

> Instructions (approximate HIT length 1-3 minutes)
>
> Find and document the OpenLibrary page of the book listed below by searching for the title and/or author here. Answering a question may reveal a follow-up question based on your response.
>
> Optionally, you may click the title below to open the Amazon.com product page for the book in a new window, allowing you to verify you have found the correct book.
>
> Some books will have an OpenLibrary page for the title, but not the specific edition. Others may not have an OpenLibrary page at all. When you have answered all questions, click submit.
>
> - Title: ${title}
> - Author: ${author}
>
> Did you locate a book with the correct title and author?
> ○ Yes
> ○ No
>
> Copy and paste the URL of the works page for the book and then click submit.
> If the URL does not include "/works/" it is the wrong kind of page. Clicking a result on a search page should load the works page. To reach it from a specific edition ("book" page), click the title in the top left corner of the page following "N editions of..."
>
> [https://openlibrary.org/works/...]
>
> Optional: Type any further comment/explanation below.
>
> Your work will be automatically rejected if you do not answer each applicable question. All other submissions will be randomly verified.
> If you have any questions or feedback or feel your work was unfairly rejected, please contact the requester. We hope to earn your trust by being both fair and timely in tasks, payment, and communication.

*Study 2: Linking to OpenLibrary – Results and Discussion*

Case 2 workers successfully found 283 potential matches (37%) for missing books in the original data. We followed up on HITs with comments and rejected submitted URLs outside the specified page types. A researcher checked every 20$^{th}$ HIT returned for accuracy during data collection and found very low rates of false matches (<1%) and false negatives (5%-10%). Checking during data collection (rather than using a simple random sample of all returned HITs) provides opportunity to save money by canceling remaining unclaimed HITs if design flaws are discovered. Consistent with case 1, the 33 workers who completed only one task in this case averaged 298 seconds, but the 50 workers who completed multiple tasks averaged only 126 seconds per task. Total cost for this case including fees was $235.

The experiment we embedded in this case illuminates how HIT design affects cost and quality. Workers presented with detailed instructions and design features spent less time per completed HIT (mean 171 seconds, S.D. 145) than those provided concise (230, S.D. 317) or minimally formatted (245, S.D. 233) instructions. However, because of the small cell sizes in this task, such differences are not significant with two-tailed T-tests; nonetheless, we take the magnitude of the differences to indicate that better instructions are likely to yield better results. Though there is a general concern that paying workers per task may lead them to rush and skim longer instructions, yielding lower quality work, we did not find that this approach compromised accuracy in our testing. Instead, work accuracy in all three groups was high and statistically indistinguishable. We speculate that fuller instructions may reduce cognitive demands on workers and thus lead to lower completion times with comparable accuracy.

*Study 3: Mental Health Websites – Overview and Methods*

Our third case study does not focus on a big data project directly. Instead, it tests the possible extent of MTurk's data augmentation capacities and directly evaluates MTurk data augmentation against a "gold standard" benchmark from a set of trained coders in an existing sociological data set. Tthis case reveals how task complexity affects MTurk results and it provides alternate methods of assessing the quality of MTurk data augmentation. In this case, we compare the performance of trained coders against MTurk workers in a study of college student mental health. The Healthy Minds Study Institutional Website Supplement (HMS-IWS) collects data on 74 topics across 8 areas related to resources, information, and the presentation of information on mental health services from college and university websites. It is, itself, adding value to a standard survey (the Healthy Minds Study) through manual data augmentation.

For three years, the HMS-IWS team, including a Ph.D. researcher and two trained graduate research assistants, have each coded relevant items from institutional websites. There is high inter-rater reliability in this manual data augmentation approach but also extensive costs and time. In this case study, we asked 40 MTurk workers to record information from one of three college or university websites. We provided workers with a brief explanation for each task (see Appendix) as well as the website link. We varied HIT construction across four categories to test how HIT organization and design affects work quality and cost. In HITs 1A and 1B, we gave workers a set of 21 items (18 yes/no and 3 open-ended) spanning four broad categories (general information, campus-specific information, information for individuals other than students, and diagnosis) and paid $1.50 for the task. In HITs 2A and 2B, we gave workers a set of 33 items that fit under a single category (services and treatment), including 30 yes/no and three open-ended questions, and paid $1.75 for the task. Finally, we varied the HITs between versions A and B, with the sole difference between versions being the addition of a paragraph in the B variants that told workers we would check accuracy and that users with too many inaccurate answers would not receive payment.

*Study 3: Mental Health Websites – Results and Discussion*

To evaluate worker accuracy, we compare results from the trained coders, which we take as a gold standard benchmark for accuracy, to results from MTurk workers. Three trained researchers first coded each of the 48 binary items for each of the three websites. The researchers agreed on 131 of the 144 total items across the three websites, and the remaining 13 items were checked again for accuracy. In contrast, MTurk workers correctly answered binary items at a rate of 63% for HIT 1A, 70% for HIT 1B, 78% for HIT 2A, and 82% for HIT 2B. Given the binary response choices, these rates are generally low. They do not improve when we use a consensus rule to aggregate MTurk responses to the same question: assuming an item's majority answer was correct would have resulted in errors for 31% of items. The accuracy difference between HIT 1A and HIT 1B is significant using an unpaired t-test ($p<0.05$), while the difference between HIT 2A and HIT 2B is not significant under the same test. The pooled difference between HITs 1 and HITs 2 is also statistically significant (p<0.001). Moreover, the pooled results show that individuals given the A variants were more likely to have a low accuracy rate than those seeing the B variants at a rate of 22% to 8%, respectively (p<0.05).

In evaluating this case, we discovered an additional finding that pertains to best practices for MTurk data augmentation. Researchers might be tempted to proxy data quality with task completion time, discarding work completed in the shortest or longest amount of time, or both. However, we found little benefit from doing so. The correlation between accuracy and completion time is 0.34, and falls slightly (to 0.29) if we remove work completed in the bottom decile of completion times. If we remove work completed in the top decile, it increases (to 0.48). Removing both changes the correlation only marginally (to 0.44). On this basis, we conclude that completion time is a weak indicator of work quality. Some who complete the task quickly may simply be good at it, while some taking the longest amounts of time may have stepped away from the computer without sacrificing work quality. Recall that MTurk workers are paid by the task, not by completion time.

Overall, results from this case show that not all data augmentation tasks can be done effectively by online crowdsourcing workers. We focused on simple yes/no questions and received a 63% accuracy rate in one HIT iteration, only marginally better than random chance. However, we can draw other important conclusions about using MTurk for data augmentation from this case: alerting workers to the possibility of payment loss from sloppy work improves accuracy (consistent with Corrigan-Gibbs et al. 2015), as does the careful ordering of work into logical groups. Finally, researchers should be careful when evaluating work accuracy, as high error rates were maintained under consensus coding and showed little relationship to completion time.

**DISCUSSION**

The use of online crowdsourcing for survey and quasi-experimental research is gaining acceptance. A series of studies that compare the results of parallel surveys and experiments using MTurk and traditional methods have evaluated online crowdsourcing with generally positive assessments (Berinsky, Huber, and Lenz 2012a; Clifford, Jewell, and Waggoner 2015; Weinberg, Freese, and McElhattan 2014). Our content analysis of published social science papers that use MTurk indicated that such evaluations have generated a set of informal norms around design and reporting for experimental and survey-style MTurk studies.

We argued that online crowdsourcing as a data augmentation platform holds unique potential to add validity and value to big data at low cost, and our content analysis suggests that researchers are beginning to use it for these purposes. However, in contrast to the emergence of norms for experimental and survey research with online crowdsourcing platforms, we found little evidence of standards for the design and reporting of data augmentation with such tools. We addressed that gap in the literature by presenting a series of three case studies designed to consider specific big data augmentation challenges, test MTurk data augmentation against known benchmarks, and improve the research community's understanding of best practices of data augmentation through online crowdsourcing.

In this section, we consider the implications of both the content analysis and our three case studies in the context of past recommendations about online crowdsourcing. We aim to provide evidence-based guidance for two types of researchers: (1) those exploring the viability of online crowdsourced data augmentation for a project, and (2) those seeking to improve the validity and value of data augmentation efforts with online crowdsourcing. Finally, we hope that future researchers, reviewers, and editors will find these considerations valuable when evaluating data quality and reporting adequacy in online crowdsourcing studies, so we offer a model reporting template in the appendix in service of this purpose.

*Strengths and Limitations of Using Online Crowdsourcing for Data Augmentation*

Our three case studies test whether and when online crowdsourcing is practical for adding validity and value to big data projects. We found that data augmentation through online crowdsourcing platforms performs best in instances like case 1, where target data are clearly defined and standardized, but it is too time-consuming, challenging, or costly to automate information recovery or for trained coders to manually recover and evaluate this information. In such tasks, workers on online crowdsourcing platforms can find and code information quickly and efficiently. The results of case 2 suggest that researchers must consider the importance of the specific output data and likely return on investment before fielding HITs. While results in this case were accurate, most books lacked a match, reducing the effective value of data augmentation through online crowdsourcing. Nonetheless, were this case focused on a larger project with tens of thousands of missing records, for instance, perhaps substantial could be gained. Case 3 looked at MTurk's potential for research beyond simple big data augmentation tasks, but it offers a more cautionary tale, wherein the non-specialized skills of online crowdsourcing workers and task completion incentives led to poor accuracy. While data augmentation through online crowdsourcing may not satisfy the complex needs of standard sociological studies such as the HMS-IWS, it can still save time and cost when used for smaller, more straightforward portions of the data collection process that would be necessary with big data augmentation.

To the extent that each of the following are true, we argue that using online crowdsourcing to augment big data should be considered more beneficial for potential cost and time savings:

1. Data collection cannot readily be automated.
2. Data can be found and/or coded by web-savvy persons without special training or knowledge.
3. Analytic needs for data are factual and do not include population estimates or comparisons with under-represented groups (minorities, individuals outside the US/India, older Americans, etc.).
4. Factual tasks can be split into smaller chunks without substantial duplication of effort.
5. Rapid results and the ability to test alternative instruments (e.g. pilot tests) are advantageous.

*Best Practices for Academic Requesters*

Given the broad range of goals, methods, and tools used by academic requesters, this section provides evidence-based guidance for maximizing the validity and value of big data augmentation using online crowdsourcing marketplaces. It assumes a researcher's goal is data augmentation, but it is also broadly applicable to surveys and experiments, with differences as noted. Once the decision has been made to use online crowdsourcing for data augmentation, a typical workflow includes three phases: design, collection, and analysis.

The design phase is most critical; it sets conditions for success in subsequent phases. Clear visual design and precise, jargon-free instructions increase worker efficiency and lower the post-collection burden on requesters to manually check data quality. Based on experimental tests in cases 2 and 3, we recommend providing comprehensive instructions and examples, but highlighting (through size, color, placement, etc.) the most important instructions for task success, as well as how work will be evaluated in payment decisions. Formative pilot studies can help to identify problems with design. If using external tools, such as pairing MTurk with survey administration platforms, it is vital to pretest HITs and ensure the correct operation of validation codes that verify external task completion. Malfunctioning codes are a common complaint on worker forums, as workers who have invested as much as an hour in a survey are unable to receive compensation. We recommend pre-testing all HITs on the requester sandbox (http://requestersandbox.mturk.com) and testing codes as part of this process.

Clear design for search or evaluation tasks faces the additional challenge of user customization and personalization. Major internet search engines often customize results based on user location and past search history. Requesters seeking to collect data that are comparable across cases should minimize variability by embedding custom search links in the directions, using non-personalized search engines such as DuckDuckGo, as we did in case study 1, and specifying how many results to use (e.g. the first 20); (K. Lewis 2015 also makes this point explicitly for other big data purposes). Search links can contain elements from the input that vary between cases, embed Boolean logic, and restrict results to specific domains.

Cases 1 and 3 demonstrated two additional principles specific to data augmentation and other factual HITs: a) iterative data collection, and b) related task grouping. Iterative data collection preferences rapid and efficient collection of a limited range of data over single-shot data collections designed to answer numerous questions. With large online crowdsourcing marketplaces, a sizable labor force is always available, and researchers can easily integrate prior task output into subsequent input. Outside of tasks requiring extensive setup or training, delaying follow-up questions to later tasks or collecting data for a sample rather than every case poses little threat to data quality. The ease of redeployment and incremental expansion generally make it better to wait when unclear whether a researcher will need a specific piece of information, preparing follow-ups as necessary.

We refer to the splitting of work into smaller and more coherent tasks as related task grouping and advocate that it improves work quality. Compared to initial single-shot versions of study 1, splitting the design into three HITs decreased cost and improved accuracy. Smart chunking lets workers self-select into tasks and not feel constrained to finish a longer task poorly to avoid sunk time. In both studies 1 and 2, a small proportion of the total number of workers completed most HITs, spending less time per HIT with at least equal accuracy. Related task grouping also avoids overpaying for work that is not completed. For example, a common application of big data augmentation through online crowdsourcing is asking workers to answer questions about a specific web link. If the link is invalid, any subsequent questions are inapplicable. If finding the initial links is also a goal, devoting a single task to identifying a suitable web address and asking subsequent workers to verify web address accuracy can save on excess pay while also providing cross-verification of the initial task's success.

Big data augmentation with online crowdsourcing is often swift and hands-off once HITs are posted, but some simple steps before, during, and immediately following HITs can improve data quality and requester reputation. Before activating a HIT, requesters can freely specify minimum worker qualifications, such as by only requesting workers with evidence of past task success or who have completed pre-tests (Leeper et al. 2015; Mason and Suri 2012 discuss tools for requesters more extensively). Requesters should also monitor their registered email during and immediately following HIT batches, as workers may contact them when they are unsure about the appropriate response, to report unclear directions or glitches, and to appeal rejections. Many circumstances, including browser malfunction, accidental user error, or common mistakes can result in rejection of ambiguous or good work, so researchers often accept all complete HITs and later remove poor quality data.

Of the phases of online crowdsourcing implementation, scholars have paid the least attention to analysis and reporting. The variety of big data, their relative lack of structure, and the priority of computer science and engineering over the social sciences in the field have contributed to inconsistent reporting. For data augmentation with online crowdsourcing tools to increase the validity and value of big data, transparency is imperative as to the procedure used to collect the data, how their integrity was verified, and relevant information on workers.

We provide a recommended reporting template in the appendix with both standard items that should be included in reporting all online crowdsourcing studies and items to use in reporting specifically for big data augmentation. We recommend researchers report on key study features, its purpose and implementation, and the exact criteria that they used to determine data quality, including at least one of several potential validity checks. Whenever possible, we suggest that both instruments and output data should be made available through public data repositories, such as the Dataverse network (www.dataverse.org) or other publicly accessible sites, such as Github repositories. In either case, standard confidentiality practices should be observed in removing unique worker numbers and other personal identifiers before publishing data, and researchers must adhere to relevant human subjects research guidelines when appropriate.

Worker compensation is a final issue that deserves discussion. Typical worker compensation among the few academic studies that report hourly pay on MTurk is $1-2 per hour, rates that prior work suggests produce reliable results (Buhrmester, Kwang, and Gosling 2011). These rates, however, are far below U.S. minimum wages and legal only because MTurk workers are

self-employed contractors not subject to minimum wage laws. Buhrmester and colleagues (2011) found that compensation was not the most commonly cited motivation for workers, but recent findings suggest many workers rely on MTurk as primary or supplemental income (Hitlin 2016; L. Irani and Silberman 2014; Litman, Robinson, and Rosenzweig 2015).[6] We worry that such low payment rates can damage the broader research community by hurting the reputation of academic researchers. A 2014 experiment (Benson, Sojourner, and Umyarov 2015) estimated that HITs from requesters with good reputations in the online review forum Turkopticon recruit workers at twice the rate of those with poor reputations (Silberman 2015; L. C. Irani and Silberman 2013). We encourage researchers who wish to estimate costs to collect a small pilot study and target average hourly compensation of at least the U.S. federal minimum wage (currently $7.25).

**Conclusion**

This paper offers data augmentation through online crowdsourcing as a means to address common concerns regarding big data in the social sciences, because doing so can add validity and value at low cost to researchers. Whereas prior work has focused on the generalizability and ethics of big data, issues of validity and value have received considerably less attention. At the same time, while many have used online crowdsourcing marketplaces such as MTurk for drawing samples, or for experimental studies, few researchers have used them for data augmentation. In this paper, we attempted to bridge these literatures. We reviewed existing practices in academic research using online crowdsourcing and considered three empirical cases where big data augmentation through crowdsourcing enhanced ongoing research or illustrated the limits of data augmentation with such tools. Based on these analyses, we provided general guidance and best practices for academic research that uses online crowdsourcing for data augmentation and a standardized reporting framework. Although we emphasized the use of online crowdsourcing for big data augmentation, many of our findings and recommendations may be of value to researchers considering online crowdsourced labor for other tasks like fielding surveys. There is substantial promise in using online crowdsourcing to free up research assistant time without the need for highly-skilled programmers, and this paper offers some first steps to formalize knowledge about the potential for using these tools.

---

[6] Litman, Robinson and Rosenzweig (2015) also find differences between US and Indian workers in both the stated importance of financial compensation and the relationship of pay rate to worker accuracy.

**Appendix 1: Reporting Template**

**How to Use**

This template provides a simple and standardized format for reporting Amazon Mechanical Turk (MTurk) results in the social sciences.[7] This version is a minimal reporting template, including a recommended set of quantities to allow reviewers and readers to evaluate the general quality of the data, its applicability, and possible limitations or problems. Because of the variety of possible uses and structures of MTurk studies, investigators are encouraged to report additional details not anticipated in this template as necessary. Items in the first section should be included in all studies reporting MTurk results. Items in the second section should be included whenever germane to the design of the study. We encourage investigators to maintain a public repository with this documentation, copies of all instruments, and (when possible) an anonymized copy of the original output.[8] The online supplement includes a sample of such a repository containing all recommended material for the case studies we review in this paper. It additionally includes (1) data and further information on the formal content analysis and (2) a suite of freely adaptable tools for Stata to help prepare raw MTurk output for analysis and public archival.

---

[7] We anticipate the template may be usable for other crowdsourcing platforms with only small modifications, but focus on MTurk as the largest and most established platform for academic use.
[8] Recommended locations for repositories are within online supplements to an article, open-access data archives, institutional repositories, or public GitHub repositories.

**Template for Reporting Social Scientific Data Collected using Amazon Mechanical Turk\***

| Recommended for all studies | |
|---|---|
| **Item** | **Description** |
| Batch | Name or signifier of batch |
| HITs | Number of HITs (unique cases in input file) |
| Workers per HIT | Number of workers assigned to complete each HIT (e.g. provided identical input) |
| Date(s) | The date(s) and time period during which the batch was collected |
| Instrument(s) + | HTML, complete description, or screen capture of instrument(s) for tasks exactly as implemented |
| Source of input data | What defines cases in the input file and where the data are originally derived from |
| Output variables | Descriptive statistics for output variables used in analysis (including missing patterns and worker demography if applicable) |
| Qualifications | List of requirements for workers to accept HITs (standard or custom) |
| Rejection criteria | Description of how decision was made to approve or reject assignments |
| Rejection rate | Proportion of submitted assignments that were rejected |
| Validation check(s) | At least one additional procedure (other than qualifications or rejection criteria) to verify data quality. Such procedures include:<br>• Consistency between multiple workers on the same HIT (inter-rater reliability)<br>• Accurate completion of items with known correct answers included in HIT<br>• Worker attention checks (questions with obvious correct answers to ensure workers are reading questions and following directions)<br>• Confirmation in later sequential HITs<br>• Consistency with another method (e.g. automated coding or trained coders) |
| **Recommended whenever applicable** | |
| **Item** | **Description** |
| Third-party tools | Name and version number (or date, if non-versioned) of any third party tools such as Qualtrics or SurveyMonkey used to administer HITs externally |
| Design features | Precise description of any contingency, experimental, or quasi-experimental design that is not clear from the instrument (often requires third-party tool) |
| Sampling methodology | Information on any sampling process, including the population being sampled, how cases were selected for inclusion, and whether the sample is with replacement |
| Weights | List of any weight or adjustment variables and their derivation |
| Panel attrition | Standard panel attrition statistics for longitudinal data collection |
| Repeat worker rate | For surveys, experiments, and other tasks collecting information about workers, the proportion of HITs completed by workers who had already completed one or more HITs in the study |
| Repeat | For tasks collecting information about workers, the proportion of demographic |

| worker consistency | responses consistent between HITs by the same worker |

\* Unless identical across batches, items should be reported for each batch of data collected using MTurk

\+ We recommend these items be included in reporting table as the URL of an online repository

Irani, Lilly C, and M. Six Silberman. 2013. "Turkopticon: Interrupting Worker Invisibility in Amazon Mechanical Turk." *Proceedings of the SIGCHI Conference on Human Factors in Computing Systems*, April, 611–20. doi:10.1145/2470654.2470742.

Jaffe, K. 2014. "Social and Natural Sciences Differ in their Research Strategies, Adapted to Work for Different Knowledge Landscapes." In *PLOS ONE* 9, e113901. doi:10.1371/journal.pone.0113901.

Kramer, Adam DI, Jamie E. Guillory, and Jeffrey T. Hancock. 2014. "Experimental Evidence of Massive-Scale Emotional Contagion through Social Networks." *Proceedings of the National Academy of Sciences* 111 (24): 8788–90.

Krupnikov, Yanna, and Adam Seth Levine. 2014. "Cross-Sample Comparisons and External Validity." *Journal of Experimental Political Science* 1 (1): 59–80.

Kuek, Siou Chew, Cecilia Maria Paradi-Guilford, Toks Fayomi, Saori Imaizumi, and Panos Ipeirotis. 2015. "The Global Opportunity in Online Outsourcing." Washington, D.C.: World Bank Group. http://documents.worldbank.org/curated/en/2015/06/24702763/global-opportunity-online-outsourcing.

Kwak, Haewoon and Jisun An. 2016. "Revealing the Hidden Patterns of News Photos: Analysis of Millions of News Photos Using GDELT and Deep Learning-based Vision APIs." Presented at the first workshop on NEws and publiC Opinion (NECO'16, www.neco.io, colocated with ICWSM'16), Cologne, Germany. https://arxiv.org/abs/1603.04531v2.

Laney, Doug. 2001. "3D Data Management: Controlling Data Volume, Velocity and Variety." *META Group Research Note* 6: 70.

Lazer, David, Ryan Kennedy, Gary King, and Alessandro Vespignani. 2014. "The Parable of Google Flu: Traps in Big Data Analysis." *Science* 343 (14 March).

Lazer, David, Alex Sandy Pentland, Lada Adamic, Sinan Aral, Albert Laszlo Barabasi, Devon Brewer, Nicholas Christakis, Noshir Contractor, James Fowler, and Myron Gutmann. 2009. "Life in the Network: The Coming Age of Computational Social Science." *Science (New York, NY)* 323 (5915): 721.

Lazer, David, and Jason Radford. 2017. "Introduction to Big Data." *Annual Review of Sociology*.

Lee, Sunshin, Mohamed Farag, Tarek Kanan, and Edward A. Fox. 2015. "Read Between the Line: A Machine Learning Approach for Disambiguating the Geo-Location of Tweets." In *Proceedings of the 15th ACM/IEEE-CS Joint Conference on Digital Libraries*, 273–274. JCDL '15. New York, NY, USA: ACM. doi:10.1145/2756406.2756971.

Leeper, Thomas J., Solomon Messing, Sean Murphy, and Jonathan Chang. 2015. *MTurkR: R Client for the MTurk Requester API* (version 0.6.17). https://cran.r-project.org/web/packages/MTurkR/index.html.

Leetaru, Kalev. 2014. "The Possibility of Global Data Sets." In *Journal of International Affairs* 68(1):215-XIII. New York.

Levitt, Gerald M. 2006. *Turk, Chess Automaton.* Jefferson, NC: McFarland & Company.

Lewis, Andrew R., Paul A. Djupe, Stephen T. Mockabee, and Joshua Su-Ya Wu. 2015. "The (Non) Religion of Mechanical Turk Workers." *Journal for the Scientific Study of Religion* 54 (2): 419–28.

Lewis, Kevin. 2015. "Studying Online Behavior: Comment on Anderson et Al. 2014." *Sociological Science* 2: 20–31.

Litman, L., J. Robinson, and C. Rosenzweig. 2015. "The Relationship between Motivation, Monetary Compensation, and Data Quality among US- and India-Based Workers on
*Enhancing Big Data in the Social Sciences with Crowdsourcing* 37